# Pressure Mode Decomposition Analysis of the Flow past a Cross-flow Oscillating Circular Cylinder

Muhammad Sufyan[1], Hamayun Farooq[2], Imran Akhtar[1] and Zafar Bangash[1]

[1]Department of Mechanical Engineering, NUST College of Electrical & Mechanical Engineering, National University of Science and Technology, Islamabad, 44000, Pakistan
[2]Centre for Advanced Studies in Pure and Applied Mathematics, Bahauddin Zakariya University, Multan 60000, Pakistan

**Abstract**

Proper orthogonal decomposition (POD) is often employed in developing reduced-order models (ROM) in fluid flows for design, control, and optimization. Contrary to the usual practice where velocity field is the focus, we apply the POD analysis on the pressure field data obtained from numerical simulations of the flow past stationary and oscillating cylinders. Since pressure mainly contributes to the hydrodynamic forces acting on the structure, we compute the pressure POD modes on the cylinder surface oscillating in lock-in and lock-out regions. These modes are then dissected into sine and cosine magnitudes to estimate their contribution in the development of pressure lift and drag decomposition coefficients, respectively. The key finding of this study is that more POD modes are required to capture the flow physics in nonsynchronous regimes as compared to synchronization case. Engineering application of this study is the development of reduced-order models for effective control techniques.

## 1. Introduction

Flow past a circular cylinder experiences many transitions with the increase in Reynolds number resulting into the von Karman vortex street [1]. The periodic shedding of vortices exert oscillatory forces on the cylinder that may lead to vortex-induced vibrations of an elastically mounted cylinder. For detail understanding of vortex shedding past a stationary and oscillating cylinders, readers are referred to Ref. [2, 3]. One of the popular approaches to analyze fluid flows is the Proper Orthogonal Decomposition (POD) [5]. It presents the given statistical ensemble with minimum deterministic modes. POD is also employed in design and optimization, flow control [6], construction of reduced-order models (ROM) [7], and closure modeling [8-11]. In literature, most of studies employed POD for ensemble velocity field data [12] while pressure field is rarely used to develop the ROM. Noack et al. [15] developed POD-based ROM and reported improved accuracy with inclusion of pressure term for the case of incompressible shear flow. However, recent studies [13,14,19] applied the POD technique for pressure field and demonstrated that application of POD on pressure field provides better insight to the flow physics. It is a fact that hydrodynamic forces depend on the pressure applied by the fluid on the structure. In this study, we further our contribution of dissecting the pressure field into modal contribution of each pressure mode for cross-flow oscillating cylinders in synchronous and nonsynchronous regimes. We decompose the surface pressure mode into its scalar sine and cosine magnitudes. The academic contribution of this research lies in the physical insight of the flow past of an oscillating cylinder using pressure modal decomposition (PMD). To the best of our knowledge, there is no work in literature that employs PMD analysis of different flow regimes; lock-in and lock-out phenomenon commonly observed in off-shore or suspended cables and tall structures. In this research, we discuss in detail the surface pressure modes, lift and drag decomposition coefficients and identify the modes that dominate the flow physics. These analysis can be quite useful in understanding the hydrodynamic forces of a moving structure, developing the reduced-order models [20] and strategies for vibration control [6,21].

## 2. Numerical Methodology

The flow past a circular cylinder is governed by the incompressible Navier-Stokes equations. In the dimensionless and strong conservative form, they can be written as

$$\frac{\partial u_i}{\partial x_i} = 0 \tag{1}$$

$$\frac{\partial u_i}{\partial t} + \frac{\partial}{\partial x_j}(u_j u_i) = -\frac{\partial p}{\partial x_i} + \frac{1}{Re}\frac{\partial^2 u_i}{\partial x_j \partial x_j} \tag{2}$$

where $i, j = 1,2$; the flow properties $u_i$ and $p$ denotes the velocity components $(u,v)$ and pressure field, respectively, along $x_i$ directions $(x,y)$. The Reynolds number is defined as $Re = U D/\nu$, where $U$ is the freestream velocity, $D$ is the cylinder diameter, and $\nu$ is the kinematic viscosity of the fluid.

We simulate the flow past a stationary and an oscillating circular cylinder at Re = 200 using an O-type grid with 193 x 257 points in the radial and circumferential directions, respectively, and the diameter of 40$D$ as shown in Fig. 1. Dirichlet and Neumann boundary conditions are applied at the in-flow and out-flow, respectively.

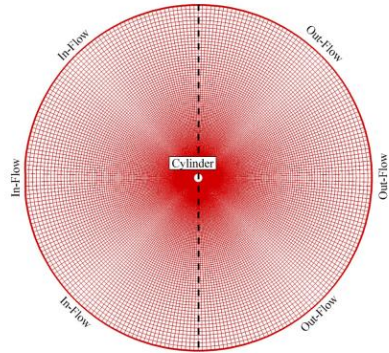

Fig. 1. O-type grid with in-flow and out-flow boundary conditions.

The governing equations are solved on a nonstaggered grid topology written in the generalized coordinates. In the present formulation, a fractional-step method is applied to advance the solution in time. A second-order central-difference scheme is used for all spatial derivatives except for convective terms where a QUICK scheme is employed. Temporal advancement is performed using semi-implicit predictor-corrector scheme, where an intermediate velocity is calculated at the predictor step and updated at the corrector step by satisfying the pressure-Poisson equation. The pressure ($p$) and viscous shear stress on the surface generate the lift and drag forces which when nondimensionlized with dynamic pressure leads to lift ($C_L$) and drag ($C_D$) coefficients of the form

$$C_L = -\int_0^{2\pi}\left(p\sin\theta - \frac{1}{Re}\omega_z \cos\theta\right)d\theta \tag{3}$$

$$C_D = -\int_0^{2\pi}\left(p\cos\theta - \frac{1}{\text{Re}}\omega_z \sin\theta\right)d\theta \quad (4)$$

where $\omega_z$ is the spanwise vorticity component.

Table 1. Comparison of numerical results.

| Ref | St | Mean $C_D$ |
|---|---|---|
| **Qua et al. [4]** | 0.1958 | 1.316 |
| **Present** | 0.19 | 1.29 |

For the flow past a stationary cylinder at Re = 200, the Strouhal number (St=$fD/U$, where $f$ is the shedding frequency) and the mean drag coefficient ($C_D$) compares well with the existing literature [3] as depicted in Table 1.

For cross-flow oscillations, an accelerated reference frame method is employed [16-18]. In a previous study [18], it was shown that for a fixed nondimensional amplitude oscillations of 0.1 ($A_e/D$) at Re=200, different frequency ratios ($f_e/f_s$ is the excitation to shedding frequency) of 0.8, 0.97, and 1.2 correspond to pre-synchronous, synchronous, and post-synchronous regions, respectively. The results obtained are compared with the experimental at Re = 200, in which Koopmann [2] identifed the lock-in region bounded by a curve in frequency-amplitude plane. From our numerical simulations, we also identify the lock-in and lock-out regions for each case. The numerical algorithm employed here is validated and verified for different flow configurations including grid and domain independent studies.

We record the pressure date with 40 snapshots per shedding cycle with 4 and 6 cycles for pre- and post-synchronous regions, respectively, covering the complete beat phenomenon.

It is important to note that only pressure forces are considered in this study since they are more dominant than the viscous forces. Qua et al. [4] reported that at Re = 100, the contribution from pressure forces accumulate to approximately 88% and 75% for lift and drag force, respectively. At Re of 200, the role of pressure forces is even more dominant.

### 3. Proper Orthogonal Decomposition

The pressure field data obtained from the snapshots is decomposed into its mean (bar) and fluctuating (prime) components as

$$p(\mathbf{x},t) = \bar{p}(\mathbf{x}) + p'(\mathbf{x},t) = \bar{p}(\mathbf{x}) + \sum_{i=1}^{M} a_i(t)\psi_i(\mathbf{x}) \quad (5)$$

where the $a_i$ are the generalized coordinates, $\psi_i$ are the pressure POD modes and $M$ is the number of POD modes employed in the Galerkin expansion. We use the method of snapshot [5] to compute the pressure POD modes [13]. From matrix of ensemble snapshots ($W$) comprising grid points as rows and time data as columns, we compute $\psi$ for which the following quantity is maximum: $\langle |\mathbf{u},\psi|^2 \rangle / \|\psi\|^2$ where $\langle \cdot \rangle$ denotes the ensemble average. We formulate a temporal-correlation function from snapshot data and transform into an eigenvalue problem. The POD modes are then computed by solving the eigenvalue problem of the form

$$GQ = Q\lambda \tag{6}$$

where $G$ is nonnegative Hermitian, $\mathbf{Q}$ and $\lambda$ are the eigenvectors and eigenvalues, respectively. The POD modes are computed as follows:

$$\psi_i = \frac{1}{\sqrt{\lambda_i}} WQ_i \tag{7}$$

From the pressure POD modes, we extract the distribution of the mode over the cylinder-surface ($S$). The surface POD is a subset of the pressure POD in full computational domain, and mathematically can be stated as $\bar{p}^s \subset \bar{p}$ and $\psi_i^s \subset \psi_i$ on $S$. We substitute Eqs. (3) and (4) into Eq. (5) and write the lift decomposition coefficients (LDC) and drag decomposition coefficients (DDC) in a Galerkin fashion neglecting the viscous terms as follows:

$$C_L^m(t) = -\int_0^{2\pi} \sum_{i=1}^M a_i(t)\psi_i \sin\theta d\theta = L_o + \sum_{i=1}^M a_i(t)L_i \tag{8}$$

$$C_D^m(t) = -\int_0^{2\pi} \sum_{i=1}^M a_i(t)\psi_i \cos\theta d\theta = D_o + \sum_{i=1}^M a_i(t)D_i \tag{9}$$

where $L_o = -\int_0^{2\pi} \bar{p}^s \sin\theta d\theta$, $L_i = -\int_0^{2\pi} \psi_i^s \sin\theta d\theta$,

$D_o = -\int_0^{2\pi} \bar{p}^s \cos\theta d\theta$, and $D_i = -\int_0^{2\pi} \psi_i^s \cos\theta d\theta$.

The subscripts "$o$" and "$i$" represents the mean and modal quantities, respectively. The $L_i$ and $D_i$ denote the LDC and DDC, respectively. We perform these pressure mode decomposition (PMD) analysis for the stationary and oscillating cylinders.

## 4. PMD Analysis - Stationary Cylinder

Using the pressure data, we compute first ten ($M=10$) pressure POD modes. The normalized eigenvalues ($\lambda$) that reveal the first few modes capture most of the system's dynamics [13]. Fig. 2 shows the contours of mean pressure mode $(\bar{p})$ and mean surface pressure mode. The mean pressure mode contour has a symmetric distribution along the horizontal axis indicating that mean coefficient of lift is zero while mean drag coefficient is non-zero. Using the pressure POD modes, we extract only the distribution of pressure modes on the cylinder surface.

Referring to Eqs (6) and (7), we investigate the contribution of $-\bar{p}^s \sin\theta$ and $-\bar{p}^s \cos\theta$ towards the generation of overall lift and drag coefficients. Superimposed on Fig. 3 is the polar distribution of the mean $(\bar{p})$ and its sine and cosine magnitudes. The radial lines can be viewed as histogram of the magnitude joined by a line. The sine and cosine magnitudes of $\bar{p}^s$ are calculated as a function of $\theta$ around the cylinder surface. These magnitudes contribute to the development of $L_o$ and $D_o$, respectively. Due to anti-symmetric distribution of $-\bar{p}^s \sin\theta$ on the upper and lower halves of the cylinder, the integration of aforementioned term over surface results in zero-mean lift. On the other hand, the distribution of $-\bar{p}^s \cos\theta$ over the upper and lower surfaces of cylinder being symmetric results in the mean drag after integrating it over the cylinder surface. Moreover, the absolute maxima of the mean surface pressure occurs at $\theta = 180°$ due to the fact that fluid becomes stagnant at this location, corresponding to the conversion of velocity head into pressure head.

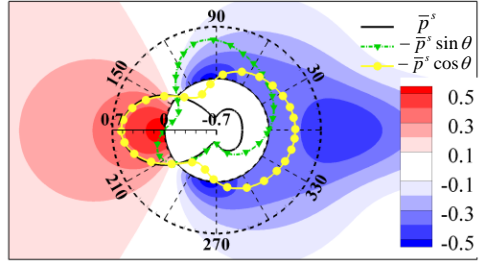

Fig. 2. Mean pressure mode overlapped with surface pressure mode.

Fig. 3 shows the first four surface pressure POD mode $(\psi_i^s)$ distribution in the polar plots depicting variation along the surface of cylinder. It is noted that the second surface mode ($i=2$) is more dominant than the first surface pressure mode ($i=1$). The dominance of the second surface pressure mode is presumably due to the dominance of the first mode downstream of the cylinder capturing the physics of vortex shedding in the wake. In Fig, 3 ($i = 2$), the sine component lies outside the cylinder resulting in a positive lift obtained after its integration over the cylinder surface. The cosine component $-\psi_1^s \cos\theta$ in Fig. 3 ($i=1$) is anti-symmetrical along the upper and lower surface of cylinder and its integration leads to almost zero contribution in the overall drag coefficient. The magnitude of $-\psi_2^s \sin\theta$ is positive and would have a major contribution towards the total lift.

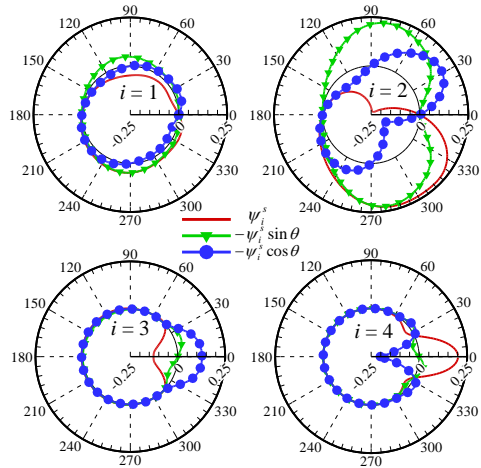

Fig. 3. First four surface pressure modes $(\psi_i^s)$ with their sine and cosine components.

Alternately, $-\psi_2^s \cos\theta$ has an anti-symmetric distribution and would result in negligible value after integration. Hence, it can be stated that the first two surface pressure modes contribute to the development of total lift. Similarly, a close look at the distribution of sine and cosine components of the third and fourth surface pressure mode reveals that the magnitude of sine components is negligible and would not have any major contribution in the total lift. On the other hand, the cosine components have absolute maxima located closer to base point therefore they would have a contribution in the generation of drag force.

We now focus on the scalar quantities of LDC and DDC. The magnitudes of the first ten coefficients are presented in Fig. 4. It is noted that in case of LDC the odd pairs are dominant, whereas the even pairs of DDC are dominant. Moreover, the first odd/even pairs of LDC/DDC are more significant than the rest of the pairs since POD provides the optimal basis functions. The sign of these coefficients depends on the temporal coefficient. The details of LDC and DDC for the stationary and their relevance with temporal coefficient can be found in previous study [14].

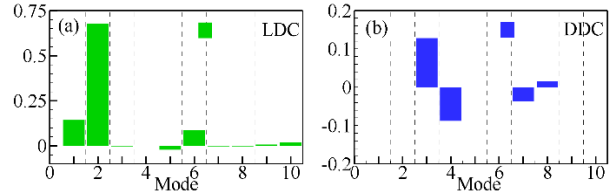

Fig. 4. LDC and DDC for first ten POD modes for the stationary cylinder.

### 5. PMD Analysis – Oscillating Cylinder

In this Section, we analyze the pressure data for the oscillating cylinder in the pre-, post-, and synchronous regimes. The POD analysis of the synchronous and nonsynchronous cases reveals that modal distribution is significantly different.

Fig. 5 presents the time histories of lift coefficient (left y-axis) and non-dimensional amplitude of excitation (right y-axis) for the cases of frequency ratios $f_e/f_s$ considered in this study. As expected, the lift coefficient for pre- and post-synchronous cases result in beat patterns with different phase angles between the excitation amplitude and the lift coefficient. On the other hand, lift coefficient of the synchronous case shows oscillations at the oscillating frequency with difference in phase angle of lift coefficient and cylinder motion as observed in the literature.

In Fig. 6, we present the first four pressure surface modes along with their sine and cosine magnitudes. It is noted that the magnitude of the first pressure surface mode appears to be insignificant on the surface as observed for the stationary cylinder. Similarly, the second mode in the post-sync case has relatively little contribution in the generation of overall lift and drag coefficient. Contrary to post-synchronous case, the second surface pressure mode from synchronous and pre-synchronous cases contribute significantly in the generation of overall forces over the structure.

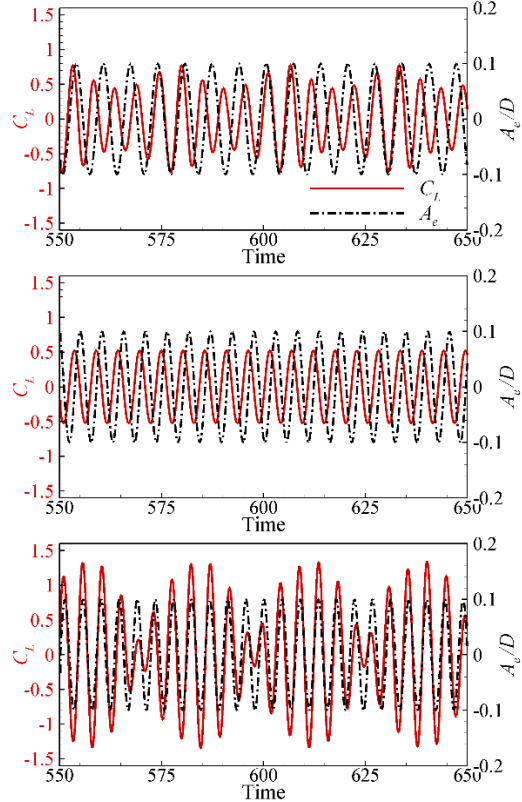

Fig. 5 Time histories of the $C_L$ and $A_e/D$ at $f_e/f_s$ = 0.8 (top), 0.97 (center) and 1.2 (bottom).

The dissection of second surface pressure mode reveals that the cosine component that contributes in the development of drag force would be negligible after integration over the cylinder surface whereas the sine components that contributes in the generation of lift force appears to have significant role both in the synchronous and pre-synchronous cases. In the post-synchronous case, the third and fourth modes appears to be more dominant and contribute in the development of lift generation upon integration of its component. The variation of drag shows that the net contribution towards the development of drag force is not significant. In the pre-synchronous case, second, third, and fourth surface pressure modes show more contribution. While the synchronous case shows qualitatively similar characteristics as that of the stationary case.

Fig. 7 presents the LDC and DDC distribution for the cases of three frequency ratios considered in this study. It is noted that the discussion presented in the previous section related pressure modes and dissection of surface modes is in line with the LDC and DDC distribution. The synchronous case have pattern similar to the case of stationary cylinder, i.e. LDC and DDC occurs in pairs.

Precisely, the first two modes of synchronous case are dominant in lift generation while second pair of modes have major contribution in the development of overall drag force. Moreover, initial modes are more dominant in generation of lift and drag forces both in the case of fixed cylinder and in the case of lock-in state. On the other hand, the lock-out cases have major contributions both from initial and higher modes in the generation of hydrodynamic forces and do not occur in pairs as in the case of a stationary cylinder. For instance, the most dominant contribution in the development of lift forces are from the second and ninth mode in the pre-synchronous case. Likewise, the tenth mode of post-synchronous case contributes noticeably in the generation of overall lift force. Relatively large magnitudes of higher modes in LDC and DDC can be attributed to nonlinear interaction of multiple frequencies in the system including the beat frequency. It can thus be concluded that nonsynchronous cases require more modes for the development of reduced-order models and in the estimation of hydrodynamic forces acting over the structure than that of the lock-in case.

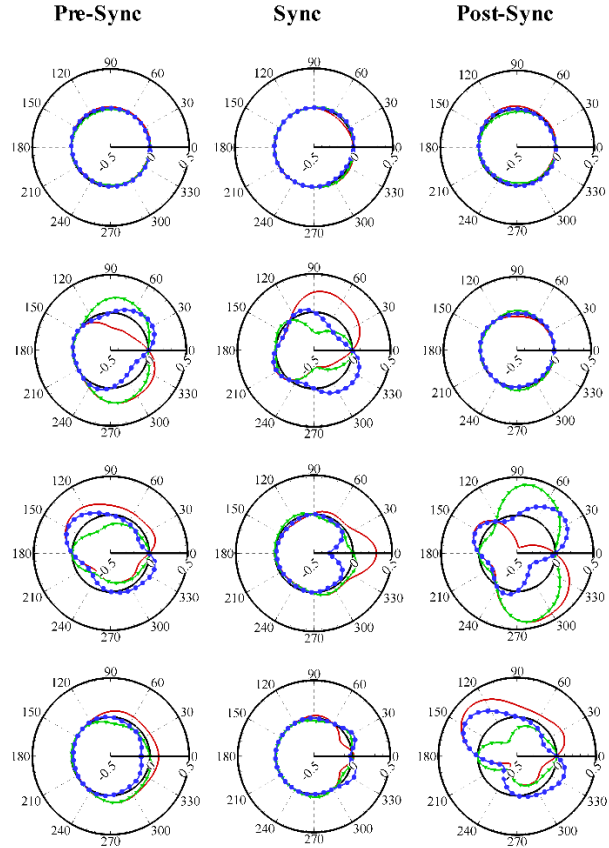

Fig. 6. Surface pressure modes 1-4 (top to bottom) for $f_e/f_s$ =0.8, 0.97, 1.2.

## 6. Conclusions

In this study, we performed numerical simulations past a stationary and an oscillating cylinder and analyzed the pressure field on the surface using the POD technique. The dissection of pressure modes reveals that in the stationary cylinder and the synchronous case of oscillating cylinder, there exist qualitative similarities. In contrast to the synchronous or the stationary case, pre- and post-synchronous cases require more modes to capture the system dynamics. We then decomposed the pressure modes into LDC and DDC to gain insight to the flow physics of synchronous and nonsynchronous flow regimes. Evidently, higher modes of LDC and DDC in nonsynchronous cases correspond to the beat phenomenon and presence of other frequencies in the system. These key findings can be beneficial in modeling hydrodynamic forces in difference flow regimes of oscillating cylinders and can further be extended to other complex geometries.

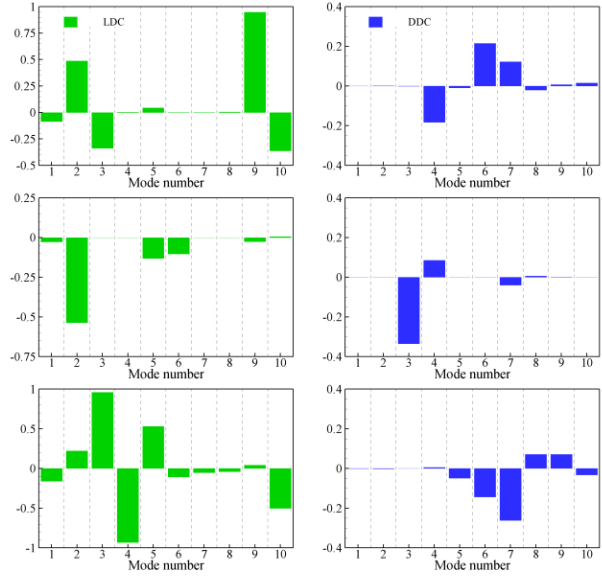

Fig. 7. LDC (left) and DDC (right) distribution of the first ten modes with frequency ratios of 0.8 (top), 0.97 (center) and 1.2 (bottom).

## Acknowledgment

This research is conducted at the Digital Pakistan Lab supported by the National Center of Big Data & Cloud Computing under Higher Education Commission, Pakistan.

## Nomenclature

| | |
|---|---|
| $A_e/D$ | : Nondimensional amplitude of oscillation |
| $a_i$ | : Generalized coordinate in Galerkin expansion |
| $D_i$ | : Drag decomposition coefficient DDC ($i^{th}$ component) |
| $f_e/f_s$ | : Excitation to shedding frequency |
| $G$ | : Nonnegative Hermitian matrix |
| $L_i$ | : Lift decomposition coefficient LDC ($i^{th}$ component) |
| $Q$ | : Eigenvectors |
| $W$ | : Snapshot matrix (rows: grid points, column: time) |
| $\psi_i$ | : Pressure POD mode ($i^{th}$ component) |
| $\psi_i^s$ | : Surface pressure POD mode ($i^{th}$ component) |
| $\theta$ | : Circumferential direction along the cylinder |
| $\lambda$ | : Eigenvalues |